# Revealing Spatial-temporal Taxi Demand Patterns after Vaccination in COVID-19 Pandemic

**Zihao Li**
Zachry Department of Civil and Environmental Engineering
Texas A&M University, College Station, Texas 77843-3136
Email: scottlzh@tamu.edu

**Cheng Zhang**
Zachry Department of Civil and Environmental Engineering
Texas A&M University, College Station, Texas 77843-3136
Email: cheng0907@tamu.edu

**Xiaoqiang Kong**
Zachry Department of Civil and Environmental Engineering
Texas A&M University, College Station, Texas 77843-3136
Email: x-kong@tti.tamu.edu

**Yunlong Zhang\***, Ph.D.
Zachry Department of Civil and Environmental Engineering
Texas A&M University, College Station, Texas 77843-3136
Email: yzhang@civil.tamu.edu

**Chaolun Ma**
Zachry Department of Civil and Environmental Engineering
Texas A&M University, College Station, Texas 77843-3136
Email: cma16@tamu.edu

\* Corresponding Author

Word Count: 6,880 words + 2 table (250 words per table) = 7,380 words

**ABSTRACT**

The COVID-19 pandemic has had an unprecedented impact on our daily lives. With the increase in vaccination rate, normalcy gradually returns, so is the taxi demand. However, the changes in the spatial-temporal taxi demand pattern and factors impacting the recovery of this demand after COVID-19 vaccination started remain unclear. With the multisource time-series data from Chicago, including pandemic severity, vaccination progress, and taxi trip volume, the recovery pattern of taxi demand is analyzed. The result reveals the taxi trip volume and average travel distance increased in most community areas in the city of Chicago after taking the COVID-19 vaccine. Taxi demand recovers relatively faster in the airport and area near the central downtown than in other areas in Chicago. Considering the asynchrony of data, the Pearson coefficient and Dynamic Time Warping (DTW) are both applied to investigate the correlations among different time series. It found that the recovery of taxi demand is not only related to the pandemic severity but strongly correlated with vaccination progress. Then, the leading and lagging relationship between vaccination progress and taxi demand is investigated by the Time Lagging Cross-Correlation method. Taxi demand starts to recover after the first-dose vaccination period started and before the second-dose vaccination period. However, it was not until mid-April that the recovery rate of taxi demand exceeded the growth of the vaccination rate.

**Keywords:** COVID-19, Vaccination, Taxi demand, Spatial-temporal pattern, Correlation analysis

## INTRODUCTION

People's daily travel patterns changed dramatically because of the outbreak of the COVID-19 pandemic at the beginning of 2020. The COVID-19 pandemic has spread worldwide. More than 188 million positive COVID cases had been confirmed, and 4 million people were dead by mid-July in 2021 (*1*). At the initial stage of the pandemic, many public places were required to be locked down to prevent the wider spread of the infection. People were permitted to travel only when necessary in many cities across the world, such as Wuhan, China (*2*), Chicago, U.S. (*3*), and Île-de-France, France (*4*). People in these locked-down cities started working and studying from home so that traffic volume dropped dramatically. Compared with the previous month, the total taxi trips in Chicago decreased by 50 percent in March, 2020 (*5*). Traffic volume had not started to increase until reopening, after the severe period of the pandemic (*6*). The difference of percentage change of traffic volumes in Illinois increased from -24% to about -8% within one month in the first initial reopening stage starting from the end of May, 2020 (*7*). Especially, the trend of traffic demand showed a noticeable increase after the time when more people were fully vaccinated (*8*). The vaccination process in the U.S. began in December 2020 and about half of the residents were fully vaccinated by mid-July in 2021 (*9*). With the vaccination rate continually rising, schools and companies in many countries declaimed their plans to go back to face-to-face study and work safely (*10, 11*).

Besides the recovery of traffic volume, the change of travel mode preferences attracts researchers' attention in the post COVID-19 period. When traveling in necessity during the pandemic, people preferred private travel mode and active transport modes, such as walking and bicycles, but not shared ones because of the fear of being infected by the coronavirus (*12*). Unlike public transit, taking a taxi is a travel mode easier for good disinfection and keeping social distancing. Some people who participated in the survey indicated that they would continue to take less bus and train trips even after the pandemic (*13*). Some respondents preferred taxi as one of the prospective transportation mode choices (*14*). Based on its advantages and previous research results, the number of passengers taking Taxis will probably recover and grow faster than that of choosing buses or subways in the post COVID-19 period. Therefore, taxi should be paid more attention to when studying traffic patterns after vaccination.

Abundant literature has studied the change of traffic mode preferences and travel patterns during the COVID-19 pandemic. Most of the current studies mentioned the lower traffic volume, less frequency of taking public transit, but relatively the same frequency of private mode trips during the pandemic compared with pre-COVID conditions. However, limited research studied the travel pattern change after vaccination and factors impacting transportation recovery. Therefore, this paper utilized the multisource data, including COVID-19 related data, vaccination data, and taxi trip data in Chicago, to quantitatively analyze the spatial-temporal trip pattern of taxi after COVID-19 vaccination started. Then, the correlation-based framework is proposed to identify the factors impacting the recovery of Taxi transportation. The findings aim to provide more information for policymakers or transportation agencies on the impact of vaccination on the recovery of Taxi trip demand.

## LITERATURE REVIEW

The COVID-19 pandemic caused worldwide adverse effects on daily travel, diminishing long-distance traveling thus mostly only necessarily trips remain. With the rapid increase of daily new COVID-19 cases and the policy of traveling restriction, office workers tended to work from home and students were required to study at home, which caused a dramatic traffic volume decline in the first stage of lockdown (*15, 16*). The sudden spread of the COVID-19 pandemic also changed the availability of travel mode and people's mode choice preferences. Travelers tended to take private transit modes for essential travel, which caused the sharp drop of traffic volume in the public transit system and other shared travel modes. Bus stops around universities even disappeared during severer lockdown periods (*17*). Compared with 70% of bus stops in other areas recovered in the new normal period, only 12% of stops around campus came back to regular use (*17*). Not only bus system was affected during the pandemic, other public transit modes still maintained a relatively low use rate within the COVID-19 pandemic period (*18*). It is not surprising that fewer public transit modes were chosen during the pandemic because of the difficulty of maintaining social distancing. At the same time, the recovery speed of public transit is also less than other modes after the severe period of the pandemic (*17*). However, some people showed their willingness to take public transit as their general mode choice with the wide use of disinfection measures in the post-lockdown period (*19, 20*). Most people still preferred not to take buses or transit after the pandemic if they had alternative transportation modes (*13*).

In the later part of 2020, the travel volume of all transportation modes slowly increased after the vaccination started. If there are enough people vaccinate in the mid-2021, the travel demand is estimated to return to pro-pandemic level in August 2024 (*8*).

For example, air transportation was estimated to fully recover in 2.4 years (*21*). However, travel patterns may not recover as anticipated in some studies. The recovery path may be influenced by various factors, such as the rising worries about financial stability among transport operators (*22*). The public transportation system will be continually affected in the following long-term because fewer operators are willing to contribute to previous jobs again. Moreover, regional disparity and income differences may also affect the recovery rate of travel patterns. Infection densities among neighborhoods need to be considered, and noticeable travel pattern differences exist in lower-income areas (*23*). A similar finding, mobility heterogeneity between the income or population density group, is also concluded in another research (*15*). Thus, more studies about factors affecting travel patterns in the recovery period should be focused on.

Although previous research significantly contributes to the differences between traffic patterns and mode choices during and before the pandemic, some gaps still exist. First, few studies investigated the recovery pattern of the transportation system in the post COVID-19 period. Second, the relationships among the pandemic severity, vaccination progress, and taxi trip volume in the recovery stage are still unknown. This study aims to mitigate the gap mentioned above by utilizing multiple datasets before and after vaccination by proposing a correlation-based framework to analyzing the factors impacting the recovery of Taxi trip demand after vaccination. The structures of this paper are as follows. Section 1 is the introduction of this paper. Then, section 2 reviews similar studies to this research and points out the research gap. Data source and data processing are described in section 3. The methodology is presented in section 4. The results of this research and details discussion are displayed in section 5. Finally, section 6 summarizes the key findings and limitations and proposes suggestions for further studies.

## DATA AND PREPROCESSING
### Data Introduction
Data in this paper is collected from Chicago Data Portal, an open-source website providing various data for the city of Chicago. This study utilized the taxi trips data, the COVID-19 data, vaccinations data, and the Chicago community area data. The taxi trip data is in 15 minutes granularity and is encrypted for privacy. It contains detailed information about the trip start timestamp, trip end timestamp, trip duration, trip distance, pickup community area, drop-off community area, and trip cost.

The COVID-19 data contains daily new and cumulative cases, death, and hospitalization in the city of Chicago. The daily vaccination counts of Chicago residents who received at least one COVID-19 vaccine dose or completed a vaccine series, including the daily and cumulative vaccination data on first dose, second dose, and total dose. The total dose is the number of people who took at least one dose. Both COVID-19 data and vaccination data are recorded by date. The 2018 Chicago community area data records the total population, area, and density of population in 77 community areas. **Table 1** shows the communities by number.

**TABLE 1. Numbering of Community Area**

| Number | Community Area | Number | Community Area | Number | Community Area |
|--------|----------------|--------|----------------|--------|----------------|
| 1 | Rogers Park | 27 | East Garfield Park | 53 | West Pullman |
| 2 | West Ridge | 28 | Near West Side | 54 | Riverdale |
| 3 | Uptown | 29 | North Lawndale | 55 | Hegewisch |
| 4 | Lincoln Square | 30 | South Lawndale | 56 | Garfield Ridge |
| 5 | North Center | 31 | Lower West Side | 57 | Archer Heights |
| 6 | Lake View | 32 | Loop | 58 | Brighton Park |
| 7 | Lincoln Park | 33 | Near South Side | 59 | McKinley Park |
| 8 | Near North Side | 34 | Armour Square | 60 | Bridgeport |
| 9 | Edison Park | 35 | Douglas | 61 | New City |
| 10 | Norwood Park | 36 | Oakland | 62 | West Elsdon |
| 11 | Jefferson Park | 37 | Fuller Park | 63 | Gage Park |
| 12 | Forest Glen | 38 | Grand Boulevard | 64 | Clearing |
| 13 | North Park | 39 | Kenwood | 65 | West Lawn |
| 14 | Albany Park | 40 | Washington Park | 66 | Chicago Lawn |
| 15 | Portage Park | 41 | Hyde Park | 67 | West Englewood |
| 16 | Irving Park | 42 | Woodlawn | 68 | Englewood |
| 17 | Dunning | 43 | South Shore | 69 | Greater Grand Crossing |
| 18 | Montclare | 44 | Chatham | 70 | Ashburn |
| 19 | Belmont Cragin | 45 | Avalon Park | 71 | Auburn Gresham |
| 20 | Hermosa | 46 | South Chicago | 72 | Beverly |
| 21 | Avondale | 47 | Burnside | 73 | Washington Heights |
| 22 | Logan Square | 48 | Calumet Heights | 74 | Mount Greenwood |
| 23 | Humboldt Park | 49 | Roseland | 75 | Morgan Park |
| 24 | West Town | 50 | Pullman | 76 | O'Hare |
| 25 | Austin | 51 | South Deering | 77 | Edgewater |
| 26 | West Garfield Park | 52 | East Side | | |

**Data Pre-processing**

Taxi trip data were cleaned by the following rules: (1) the start timestamp and end timestamp should exist; (2) the timestamp should be between 2018 and 2021; (3) the start date and time should be earlier than the end date and time; (4) travel duration of each trip should be larger than 60 seconds; (5) travel distance should be longer than 0.5 miles; (6) the pickup and drop-off community area should exist. After the cleaning, the data size has dropped from 151,853,944 to 127,970,168. Then, the 15-min taxi trip data are aggregated into daily data by the same date, pickup, and drop-off community area. At the same time, the COVID-19 data and vaccination data are also cleaned and aggregated.

The progress of COVID-19 testing and vaccination is impacted by the weekend (*24*). Besides, the taxi trip volume is also impacted by the weekend since the commuting flow decreased largely on the weekend. Therefore, a 7-day rolling average smooths out the dips on weekends is used. All analysis is based on the 7-day rolling average data.

**Data Description**

The trip data was recorded from December 31[th], 2018, to May 31[st], 2021. The COVID-19 related data is recorded from March 1[st], 2020 to June 21[th], 2021 and the vaccination data is from December 15[th] 2020 to June 21[st], 2021. This paper aims to explore the taxi trip pattern after vaccination and the factors affecting taxi trip demand recovery, so taxi trip data is divided into two parts: before and after vaccination started.

The period before vaccination is from the governor's announcement of a disaster proclamation for the state of Illinois to the start of vaccination, which is from March 9[th], 2020, to December 15[th], 2020. The period after vaccination is from December 16[th], 2020, to May 31[th], 2021. **Table 2** summarizes the selected time series data before and after the vaccination started, and the value in the table are rounded to the nearest whole number.

**TABLE 2. Descriptive Statistics of the Selected Data before and after Vaccination**

| Description | Mean | Std. | Min | Median | Max |
|---|---|---|---|---|---|
| **Before vaccination (281 days)** | | | | | |
| **03/09/2020 - 12/15/2020** | | | | | |
| daily taxi trip | 4094 | 215 | 1147 | 3853 | 30680 |
| daily new confirmed cases | 678 | 36 | 8 | 366 | 2482 |
| daily new hospitalizations | 71 | 3 | 6 | 45 | 175 |
| daily new death | 14 | 1 | 0 | 7 | 48 |
| cumulative cases | 67815 | 2795 | 15 | 60574 | 191308 |
| cumulative hospitalizations | 10706 | 289 | 35 | 11422 | 20145 |
| cumulative death | 2337 | 65 | 0 | 2784 | 3988 |
| **After vaccination (167 days)** | | | | | |
| **12/16/2020 - 05/31/2021** | | | | | |
| daily taxi trip | 5590 | 1281 | 3610 | 5299 | 8085 |
| daily new confirmed cases | 552 | 307 | 92 | 503 | 1326 |
| daily new hospitalizations | 49 | 19 | 18 | 49 | 93 |
| daily new death | 9 | 5 | 4 | 7 | 24 |
| cumulative cases | 247925 | 24834 | 192694 | 247460 | 283464 |
| cumulative hospitalizations | 24805 | 2190 | 20235 | 24775 | 28317 |
| cumulative death | 4967 | 386 | 4014 | 5052 | 5512 |
| daily new first vaccination | 8950 | 4855 | 118 | 8559 | 18264 |
| daily new second vaccination | 6983 | 4914 | 0 | 7000 | 17201 |
| daily new total vaccination | 15933 | 8633 | 118 | 15769 | 32056 |
| cumulative first vaccination | 681144 | 521554 | 129 | 570217 | 1493976 |
| cumulative second vaccination | 412061 | 399976 | 0 | 278541 | 1162109 |
| cumulative total vaccination | 1093205 | 916503 | 129 | 848758 | 2656085 |

## METHODOLOGY

This paper proposes two steps to analyze taxi trip volume, COVID-19 and vaccination data, to gain a deeper insight into the recovery of taxi trip demand after the vaccination started, as shown in **Figure 1**. As COVID-19 has a significant impact on transportation, it is common sense that the trip volume hugely decreased during the COVID-19 pandemic (*25*). As more and more people are vaccinated, taxi, an effective alternatives transportation mode to transit, may gradually recover (*26*). Therefore, it is essential to quantitatively understand the taxi trip pattern change after the vaccination at the city and region levels. First, the taxi trip volume, COVID-19, and vaccination data in Chicago are analyzed to help build a macroscopic understanding. Then the normalized taxi trip volume and average travel distance before and after the vaccination of all community areas in Chicago are analyzed to explore the regional taxi trip pattern change.

Next, the Pearson coefficient is applied to analyze the correlation among pandemic severity, vaccination progress and taxi trip volume. Dynamic Time Warping (DTW) is used to verify the correlation result from the Pearson coefficient when the data is unsynchronized. Analyzing the result from two methods, the factors most correlated with the recovery of taxi trip demand are explored. Then, the leading and lagging relationship between the factors and taxi trip volume are analyzed to further understand the recovery pattern of taxi trip demand.

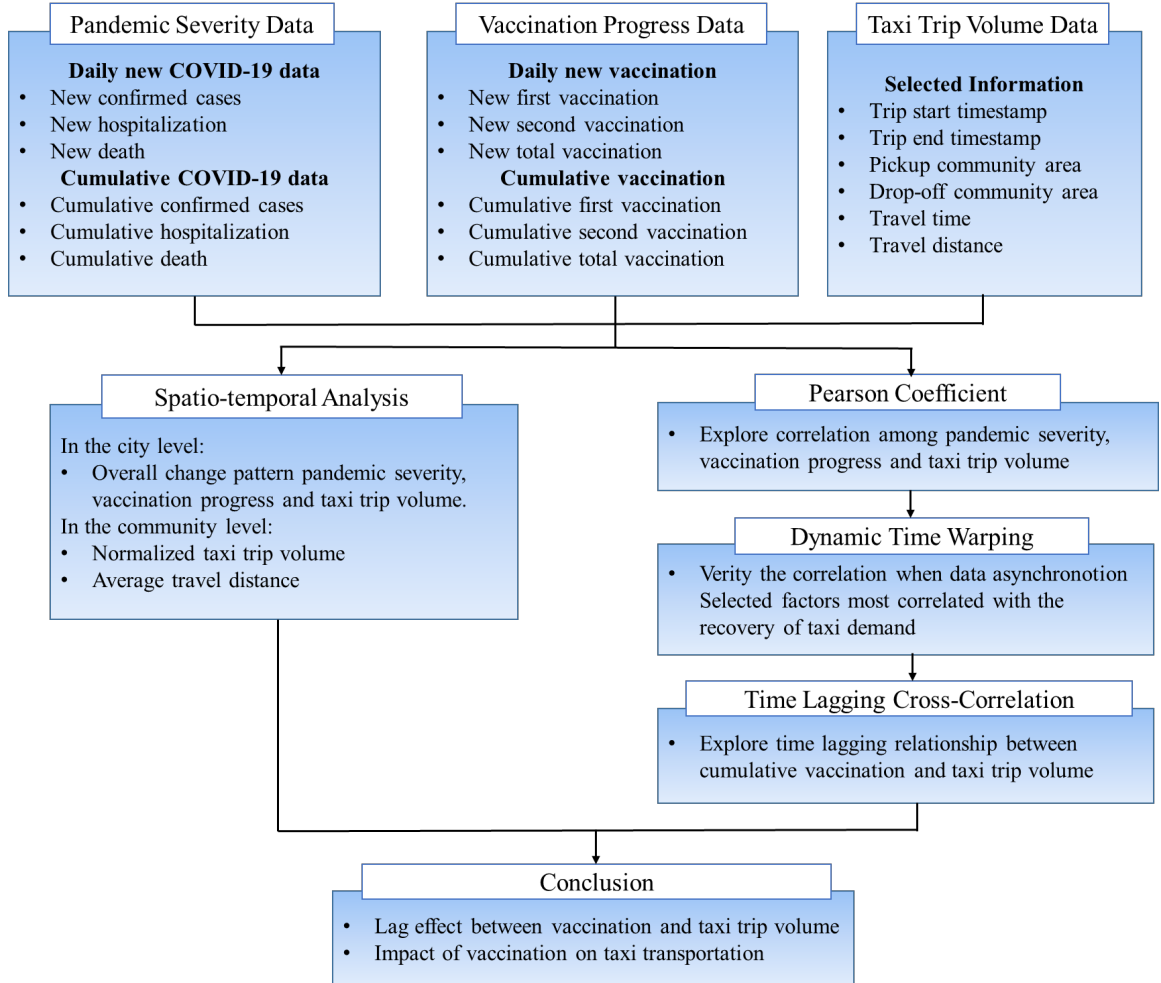

**Figure 1. Flow chart of methodology of this paper**

**Correlation analysis**

*Pearson correlation coefficient*

The Pearson coefficient (*27*) is usually used to measure the degree of correlation between two variables, whose value is between -1 and 1. The larger the value is, the significant the correlation between the two variables is. The negative and positive respectively represent the variables are negatively and positively correlated with each other. As shown in **Equation 1**, the Pearson coefficient is defined as the quotient of covariance and standard deviation between two variables and it could calculate two variables with very different values without normalization (*27*).

$$r = \frac{\sum_{i=1}^{n}(X_i - \bar{X})(Y_i - \bar{Y})}{\sqrt{\sum_{i=1}^{n}(X_i - \bar{X})^2}\sqrt{\sum_{i=1}^{n}(Y_i - \bar{Y})^2}}$$

(1)

Where,

$X_i$, $Y_i$ is the $i^{th}$ data in the $X$,$Y$ variable,

$\bar{X}$, $\bar{Y}$ is the arithmetic means of the $X$, $Y$ variable.

*Dynamic Time Warping (DTW)*

The lengths of the two similar time series may not be equal, so that the similarity between two time-series data cannot be effectively calculated by using the point-to-point Euclidean distance. When two time series have very similar waveform shapes, but the data are not aligned on the timeline, one (or both) of the time-series data is needed to warp under the timeline to achieve better synchronization and to obtain the similarity. DTW is a method to achieve the warping distortion and calculates the similarity between the two data by extending or shortening the time series. Therefore, DTW could overcome the restriction that the pair of Euclidean distance points must correspond to calculate unsynchronized data. The taxi trip volume, COVID-19 related and vaccination data may not be synchronous, therefore DTW is needed to further analyze the correlation among taxi trip volume, pandemic severity and vaccination progress. Besides, if there is a great value difference between two time-series data, normalization is needed because the distance of DTW in this paper is defined by Euclidean distance. The steps of the algorithm are as following:

Suppose there are two time series $\boldsymbol{Q} = (q_1, q_2, \ldots, q_i, \ldots, q_n)$ and $\boldsymbol{C} = (c_1, c_2, \ldots, c_j, \ldots, c_m)$, whose lengths are $n$ and $m$. To synchronize the two time series, the $n$ x $m$ matrix is built and the element $(i, j)$ represents the distance between $q_i$ and $c_j$. Then, dynamic programming is applied for searching the shortest path, which means to warp the time series best. The warping path is defined as $\boldsymbol{W}$ in **Equation 2**.

$$\boldsymbol{W} = (w_1, w_2, \ldots, w_k, \ldots, w_K) \qquad \text{Where,} \quad \max(m, n) \leq K < m + n - 1 \qquad (2)$$

The constrained optimization problem to find the best warping path is written as follows:

$$\text{min} \qquad \text{Dist}(\boldsymbol{W}) = \sum_{k=1}^{k=K} \text{Dist}(w_i, w_j)$$

$$\text{subject to} \quad w_1 = (1,1), \; w_K = (m, n) \qquad\qquad\qquad (3)$$

$$w_k = (i, j), \; w_{k+1} = (i', j') \quad i \leq i' \leq i+1, \; j \leq j' \leq j+1$$

The warping path could minimize the distance between two time series. Starting from the origin point, the $\boldsymbol{Q}$ and $\boldsymbol{C}$ are matched, and the distance calculated by all previous points is accumulated. After reaching the destination point, the cumulative distance, as shown **in Equation 4**, is the similarity between $\boldsymbol{Q}$ and $\boldsymbol{C}$. The smaller the value is, the greater the similarity between two time series and the more significant correlation between two datasets.

$$D(i, j) = Dist(i, j) + \min[D(i-1, j), D(i, j-1), D(i-1, j-1)] \qquad (4)$$

**Lagging correlation between different time-series data**

Pearson coefficient and DTW could evaluate the correlation, but neither can identify directionality between two time series, for example, the leader-follower relationship. Time Lagging Cross-Correlation (TLCC) is involved in assessing the directional relationship between two time series**,** in which the leading data initializes a response and the following data repeats it. TLCC is measured by gradually moving a time series vector and repeatedly calculating the correlation between the two time series. If the maximum correlation is in the center (offset=0), it means that the two time series have the most significant correlation at this time and they are synchronous. However, if one time series is leading or lagging the other, the maximum correlation is on the either side of the center. The calculation of TLCC is as follows:

$$r_l = \frac{\sum_{t=1}^{n}(x_{t-l}-\overline{x})(y_t-\overline{y})}{\sqrt{\sum_{t=1}^{n}(x_{t-l}-\overline{x})^2(y_t-\overline{y})^2}} \quad (5)$$

Where,

$x, y$ are two time series, and $y$ is the benchmark,

$n$ is the number of a dataset,

$l$ is the offset, representing the leading and lagging relationship, such as $l = 0$ indicates synchronization, $l < 0$ indicates leading, and $l > 0$ indicates lagging.

TLCC is used to further detect which data changes cause another data to change, and to determine the sequence of various events (*28*). First, Taxi trip volume is offset to continuously analyze the degree of deviation between it and significantly correlated time series selected from Pearson coefficient and DTW. Therefore, the maximum cross-correlation coefficient under the offset is obtained. The offset indicates the chronological order of the selected time series and the recovery of taxi trip demand. Meanwhile, the maximum cross-correlation coefficient presents whether the selected time series and the recovery of taxi trip demand show similar trends.

## RESULTS AND DISCUSSION

The exploratory analysis of the taxi trip data, pandemic severity, and vaccination progress is first presented in this section. The spatio-temporal pattern of taxi trips before and after vaccination is then explored, and the factors affecting the recovery of taxi trip demand are further investigated.

### Spatio-temporal pattern of taxi trips after vaccination
*Macroscopic level*

The blue and red lines in **Figure 2** represent daily and 7-day average taxi trip volumes, respectively. The 7-day average taxi trip volume in the whole city was from 30,000 to 50,000 trips per day before the COVID-19 pandemic. However, a considerable drop in taxi volume appeared in March 2020, the start of the COVID-19 pandemic outbreak in the U.S. The daily taxi trip volume in the lockdown period decreased to less than 10,000 trips since then. Although the COVID-19 vaccination had launched in December 2020, the recovery process of taxi trips was slow initially.

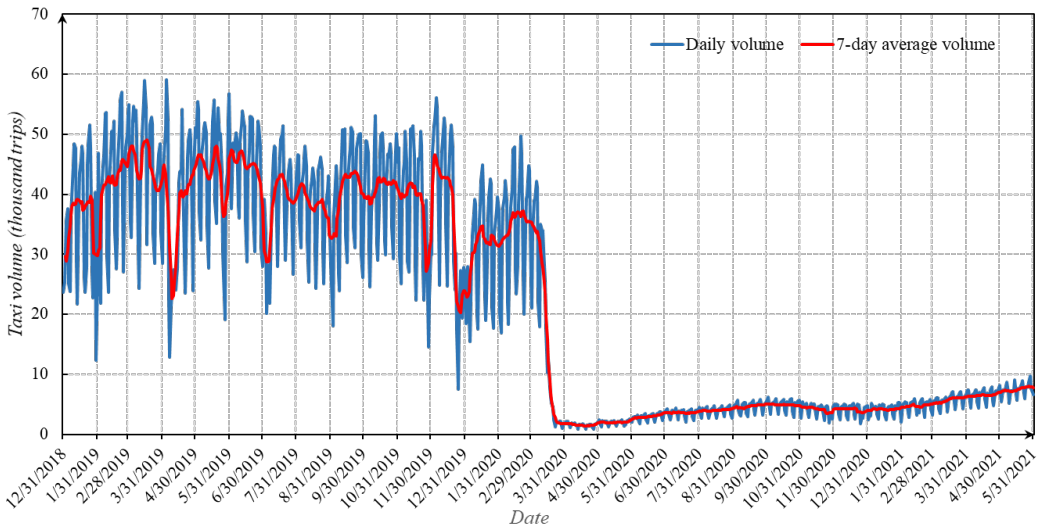

**Figure 2. Daily and 7-day average taxi volume in the COVID-19 pandemic**

**Figure 3(a)** indicates the trends of daily new COVID-19 cases, hospitalization and death, and the number of vaccinated people. There are two peaks of daily cases in 2020, and one appeared in 2021. The peak of daily new confirmed cases near November 2020 is probably related to the presidential election. Another peak occurred around April 2021, after the peak of the daily vaccinated people with the first dose. Increasing the first-dose vaccination rate boosted people's confidence in going out and encouraged them to rush back to normalcy. People who were not vaccinated and did not keep social distancing became the main population for the new peak of daily new confirmed cases in 2021. However, the number of daily new cases dropped again as more people become fully vaccinated. Besides, no matter before or after vaccination started, hospitalization and death have a similar trend with the curve of daily new cases. **Figure 3(b)** shows the trend of cumulative COVID-19 related data and the number of vaccinated people. The number of vaccinated people is much higher than the number of confirmed cases. The percentage of the adult population who have received at least one dose of the COVID-19 vaccine is around 52.5% at the end of May 2021 in Chicago. However, two threats impeded the recovery of taxi demand. One is the herb immunity has not been acheived (*29*), the other is potential virus mutation. Traveler's concern for these two threats reflects in the taxi trip demand that the taxi trip start to recover with the vaccination rate rising, but the pace of recovery is not as fast as it was expected.

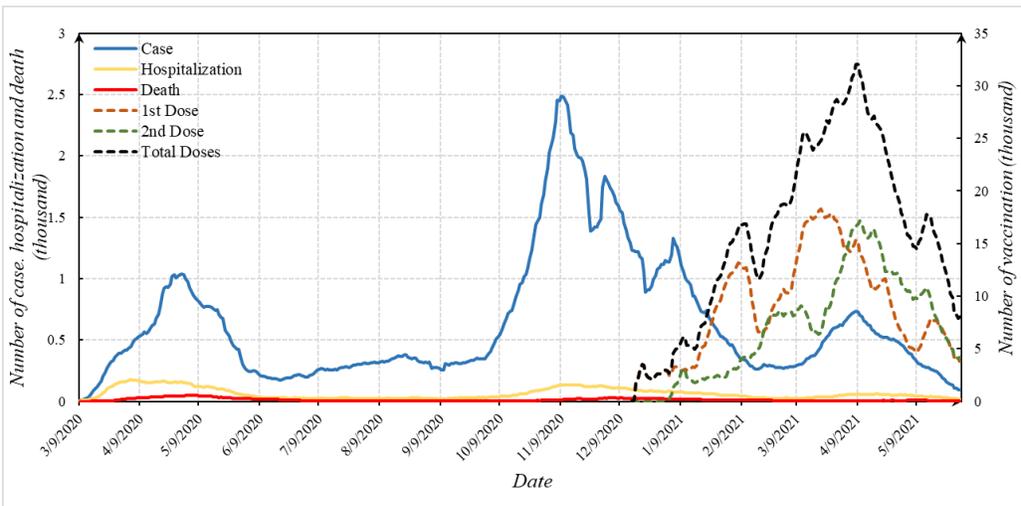

**(a)**

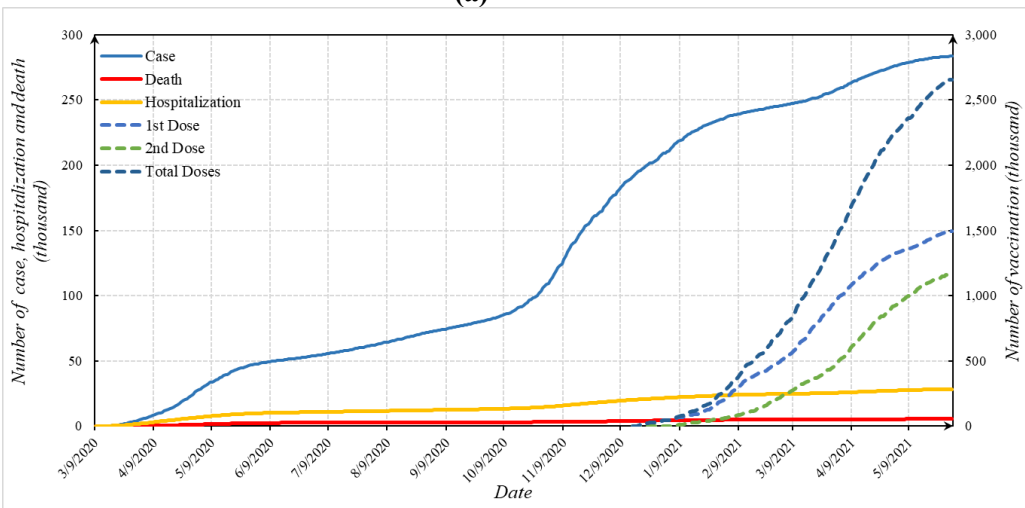

**(b)**

**Figure 3. Daily new and cumulative COVID-19 related data (a), and vaccination data (b)**

*Mesoscopic level*

The taxi trip volume distributed unevenly among community areas in Chicago, according to **Figure 4**. High volume areas are mostly located in central downtown areas and O'Hare airport before and after the vaccination. After vaccination started, community areas with the most apparent increase of the taxi trip volume are along the Michigan Lake in central and northern districts of Chicago and the airport area. The highest taxi trip volume areas in central areas are No. 8 (Near Northside) and No. 32 (Loop). People feel safe to travel when they or people around them took the vaccine, and more citizens are willing to take a taxi as before. **Figure 5** intends to investigate the relationship between the taxi trip volume before and after vaccination among 77 community areas in Chicago. Because of the different population sizes in these areas, taxi trip volume data in the different areas have been normalized by population. The red points in **Figure 5** represent taxi trip volume in a specific area divided by its population size, and the unit is the number of trips per thousand residents. It turns out that taxi trip volume before and after vaccination can be described with a linear relationship with an R square value equal to 0.9 and a slope is around 0.63. If the slope is equal to one, it illustrates that the normalized taxi trip before and after vaccination is similar. When the slope is less than one, it suggests that the normalized taxi trip after vaccination is larger than before vaccination in most community areas. Moreover, when the normalized taxi trip volume after vaccination started is small, it strongly correlates with that before vaccination started. On the other hand, the larger the normalized taxi trip volume before vaccination started, the harder to estimate the recovery of taxi demand after vaccination.

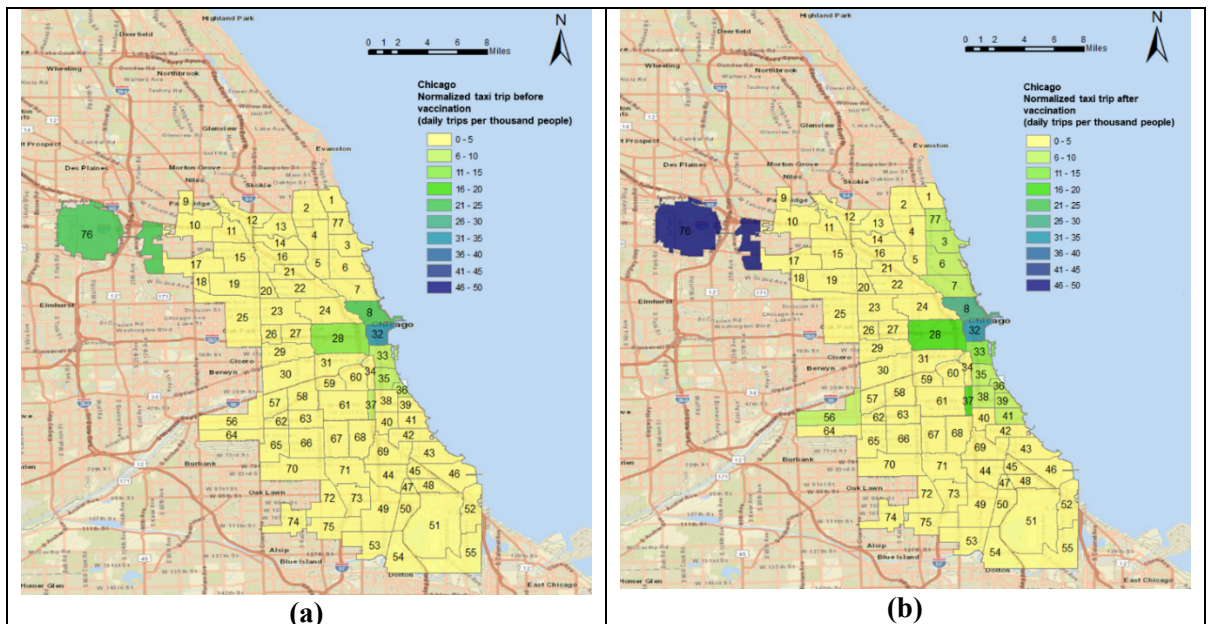

**Figure 4. Normalized taxi trip volume in each community area before (a) and after (b) vaccination**

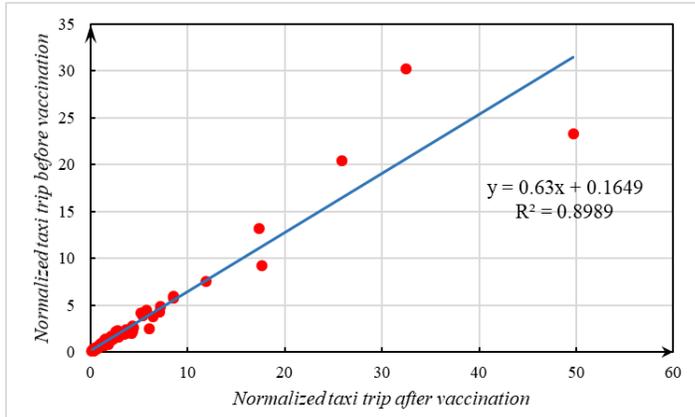

**Figure 5. Relationship between taxi trip before and after vaccination**

The green and red circles in **Figure 6** present the average travel distance before and after vaccination. If the average travel distance after vaccination is longer than before vaccination in the community area, the green circle with red periphery is shown in the area. On the contrary, the red circle with a green periphery is shown. The wider the periphery, the difference in travel distance is larger. The threshold to determine whether the difference in travel distance is significant or not is two miles. If the difference is not significant, only a green circle is drawn in the area. Average travel distance in most community areas increases after vaccination. This phenomenon is because of the stay-at-home policy during the pandemic and the fear of pandemic before vaccination, people may only travel when necessary. After being vaccinated, people are inclined to expand their activity circle and travel for multiple purposes. Only a few areas have less average travel distance after vaccination, including No. 8, 52 and 64. The reason might be the lockdown of surrounding markets in those areas during the pandemic, so residents have to travel longer distances for shopping during the pandemic. After vaccination, all kinds of business activity are reopened and people could purchase the necessities or groceries near where they live.

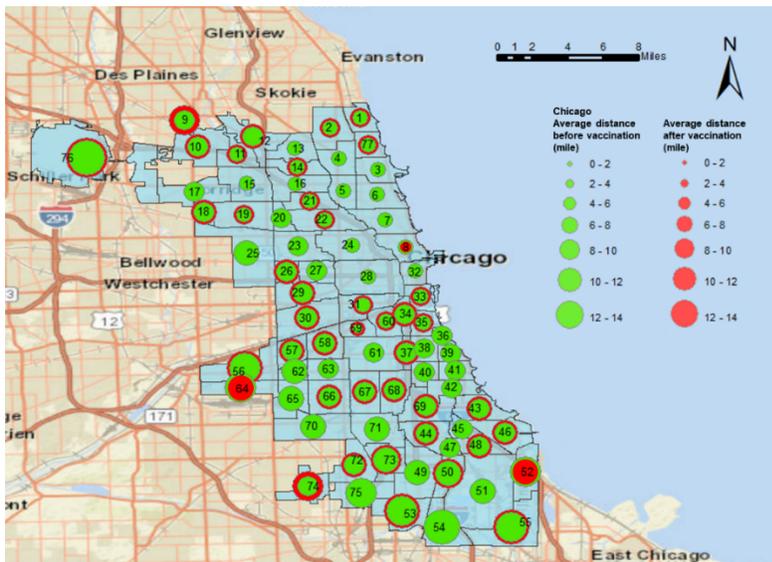

**Figure 5. Average travel distance in each community area before and after vaccination**

**Correlation among taxi trip volume, COVID-19 related and vaccination data**
*Result of correlation analysis*

The result of the Pearson coefficient analysis among different time-series data from December 15th, 2020 to May 31st 2021 is shown in **Figure 7**. The horizontal and vertical axes are the thirteen selected time-series data, including daily new and cumulative cases, hospitalization and death data, daily new and cumulative first dose, second dose and total dose vaccination, and taxi trip volume. The number in the picture is the Pearson coefficient between two time-series data. Blue indicates a positive correlation, and red indicates a negative correlation. Moreover, the darker the color, the higher the correlation. The higher the correlation means two time series data are more correlated.

The daily new confirmed COVID-19 cases, hospitalization, and death negatively correlate with the daily new vaccinated people. The number of death is the most (negatively) correlated with daily new vaccinated people among the daily new COVID data. The result demonstrates the effect of the vaccination on suppressing mortality and mitigating the severe illness rate of COVID-19 (*30–32*). On the other hand, the negative correlation between daily new COVID-19 data and taxi trip volume illustrates that the trend of taxi trips volume is the opposite of daily new confirmed cases, hospitalization, and death. The daily new death has a stronger correlation with taxi trip volume than the daily new case and hospitalization. In plain word, when the daily new COVID-19 data decrease, the taxi trip volume tends to increase, and the tendency is more evident between daily new death and taxi trip volume. The fewer people died because of COVID-19 daily, the more people chose to travel by taxi. A stronger correlation between cumulative COVID-19 data and taxi trip volume presents that, at the end of the COVID-19 pandemic, the recovery trend of taxi trip volume is in parallel with the cumulative COVID-19 related data. By comparing the correlation between daily new vaccination and taxi trip volume with the correlation between cumulative vaccination and taxi trip volume, it is found that the recovery of taxi trips is more related to the cumulative vaccination rather than daily new vaccination. When the cumulative vaccination population is continuously growing, people are inclined to take taxis more due to the increasing confidence in the vaccination rate.

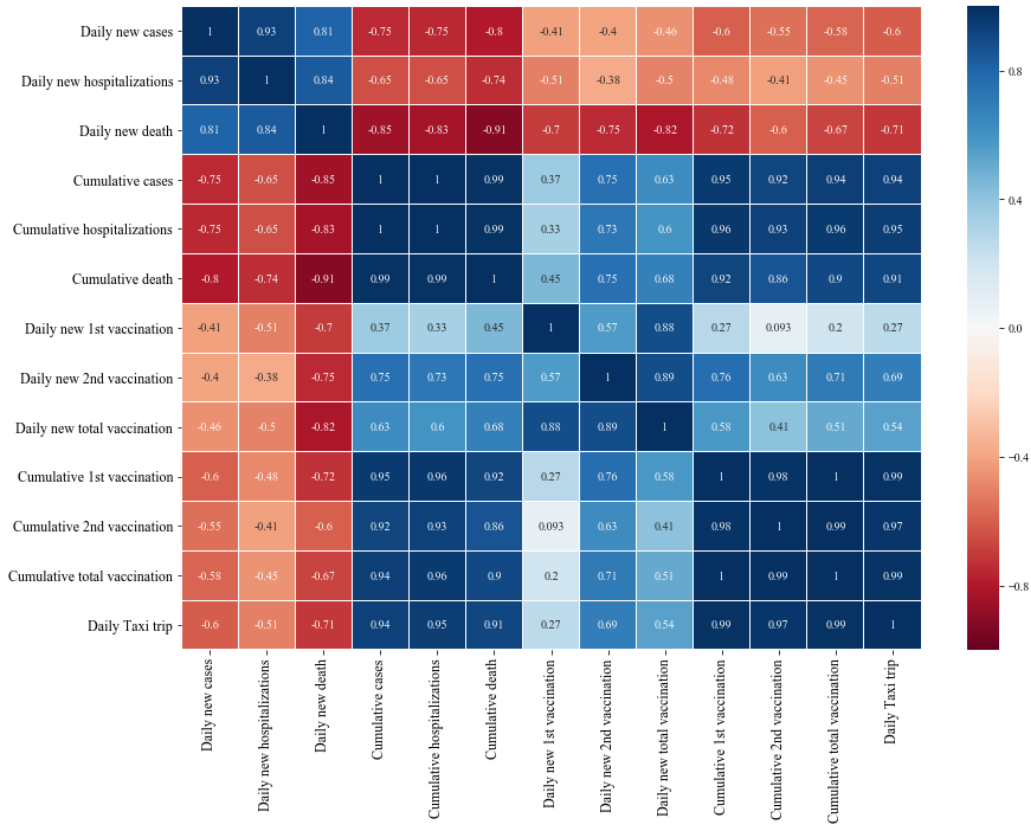

**Figure7. Pearson coefficient among taxi trip volume, COVID-19 related and vaccination data**

Since the Pearson coefficient can only handle the synchronized time series, the DTW is needed to verify the correlation when data is unsynchronized. Each plot in **Figure 8** indicates the correlation between pandemic severity and taxi trip volume and the correlation between vaccination progress and taxi trip volume is plotted in **Figure 9**. The correlation measurement is reported on the top of each plot. The smaller the distance, the stronger the correlation. The background of the picture is the distance metrics between each point in the time series. The darker the color, the smaller the distance between two points. Besides, the white line is the path with the minimum distance between two time series.

The same result as the Pearson coefficient could be obtained by comparing the minimum distance between different time series and taxi trip volume. **Figure 8 (a)** to **(c)** and **Figure 8 (d)** to **(f)** accordingly show the minimum distance of daily new COVID-19 data and cumulative COVID-19 data to taxi trip volume. Besides, **Figure 9 (a)** to **(c)** and **Figure 9 (d)** to **(f)** presents the minimum distance of daily new vaccination data and cumulative vaccination data to taxi trip volume. The average minimum distance of daily new COVID-19 data (216.05) and cumulative COVID-19 data (20.73) are greater than daily new (83.15) and cumulative (11.86) vaccination data. Therefore, recovery of taxi demand is more related to the vaccination progress than the pandemic severity.

As shown in **Figure 9**, the minimum distance between taxi trip volume and cumulative vaccination is smaller than that between taxi trip volume and daily new vaccination. Therefore, cumulative vaccination, including the cumulative first dose, second dose and total vaccination, could be better indicators of the recovery of taxi demand. Moreover, cumulative vaccination should be the most significant factor impacting the recovery of taxi demand from the results of Pearson coefficient and DTW.

However, Comparing **Figure 8 (b)** with **(a)** and **(c)**, and **Figure 8 (e)** with **(d)** and **(f)**, the hospitalization is the most correlated factors to taxi trip volume among COVID data, which is different from the result of Pearson Coefficient. In the Pearson Coefficient, the most related factor among the COVID data is death. the difference may cause by the significant association between hospitalization and death caused by COVID-19 (*33*).

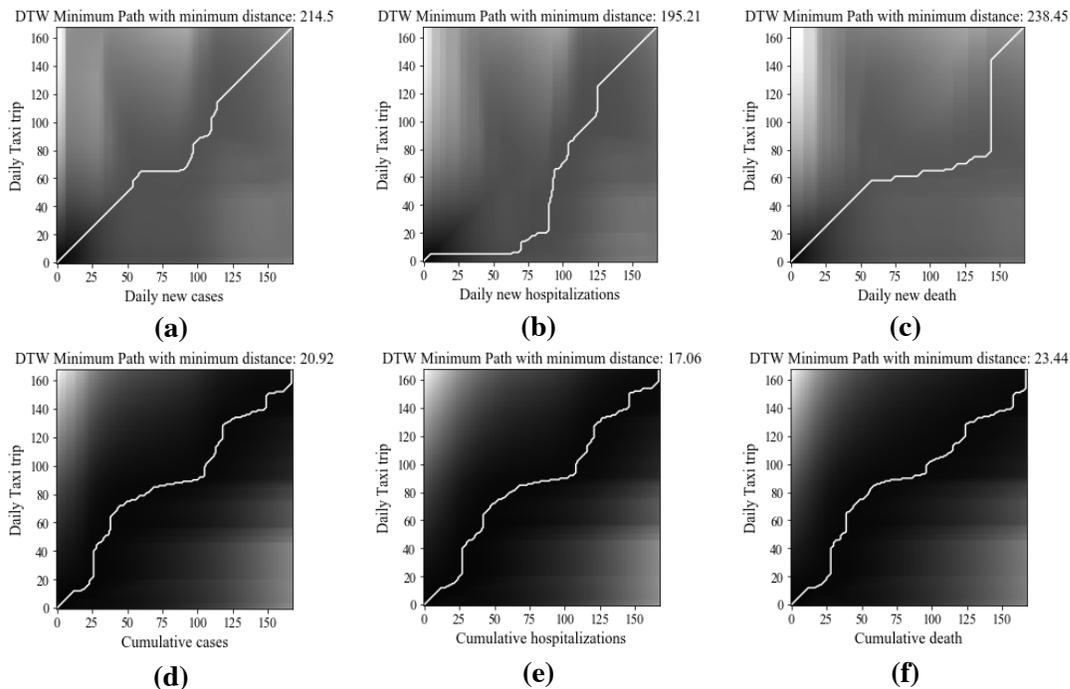

**Figure 8. DTW result between taxi trip volume and COVID-19 related data**

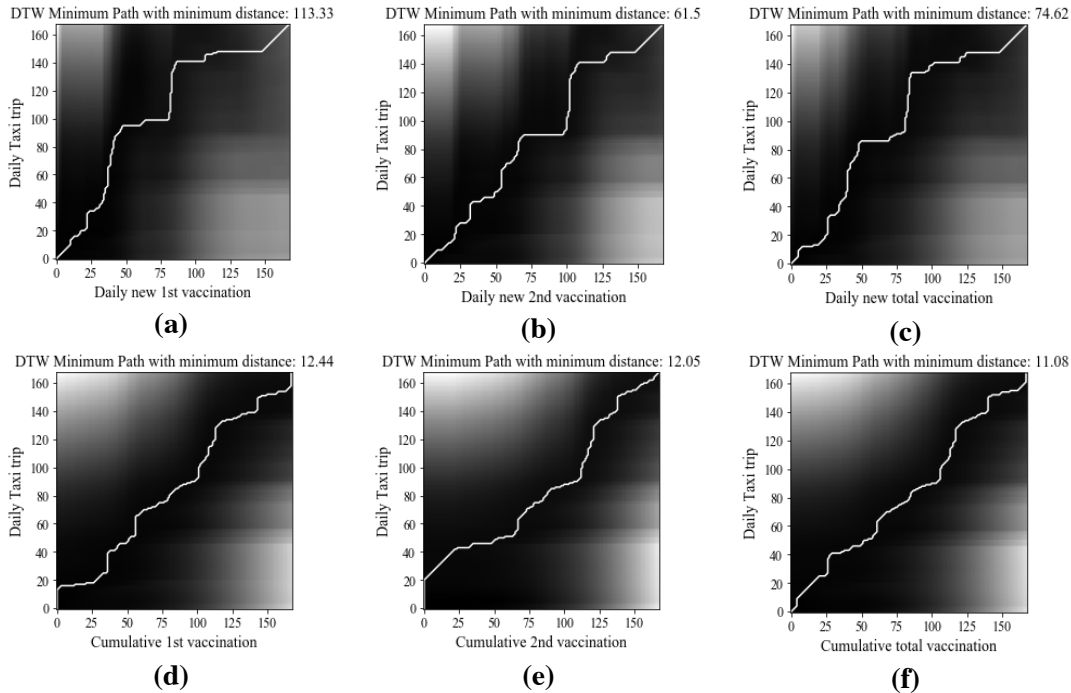

**(a)** **(b)** **(c)**

**(d)** **(e)** **(f)**

**Figure 9. DTW result between taxi trip volume and vaccination data**

*Result of lagging correlation analysis*

Correlation analysis of factors related to taxi recovery has been explored, but the leading and lagging relationship between different time series is unknown. Therefore, the most relevant factors- cumulative vaccination, including cumulative first dose, second dose and total dose, are selected to examine the leading and lagging relationships with taxi trip volume.

The black line in **Figure 10** presents the centerline as a reference and the red line indicates where the time lagging cross-correlation is the greatest. The offset in the title of each picture shows the leading and lagging relationship. If the value is larger than zero, it reveals that the cumulative vaccination number leads the taxi trip volume. Otherwise, the taxi trip volume leads the cumulative vaccination number. The smaller the offset, the faster the lagging signal is affected by the leading signal. The frame of offset represents the number of days leading or lagging. The difference in the offset could reflect the speed of change. In **Figure 10(a)**, the cumulative number of first dose vaccination leads the recovery of taxi trip volumes with two frames offset, which means taxi trip volume increases after the cumulative number of vaccinated people with the first dose two days. From **Figures 2** and **3**, at the beginning of vaccination, the increased taxi trip volume is greater than people who took the first vaccination, but the growth rate of taxi trips is lower than that of cumulative first vaccination. With the increase of people vaccinated, the recovery rate of taxi demand exceeds the growth of vaccination rate until mid-April 2021. **Figures 10(b)** indicate taxi trip volume leads the number of people who took the second dose with 25 frames offset. In other words, the recovery of taxi demand is 25 days earlier than people who took the second dose vaccine. The phenomenon is because of the recommendation from the Centers for Disease Control and Prevention (CDC) that the second dose vaccine should be administered 3-week or 4-week after the first dose. Therefore, the result is consistent with the result from **Figure 10 (a)**. The small negative offset in **Figures 10(c)** is because the total vaccination counts the people who took at least one dose. Hence, the offset is close to the average offset of the cumulative first vaccination and second vaccination.

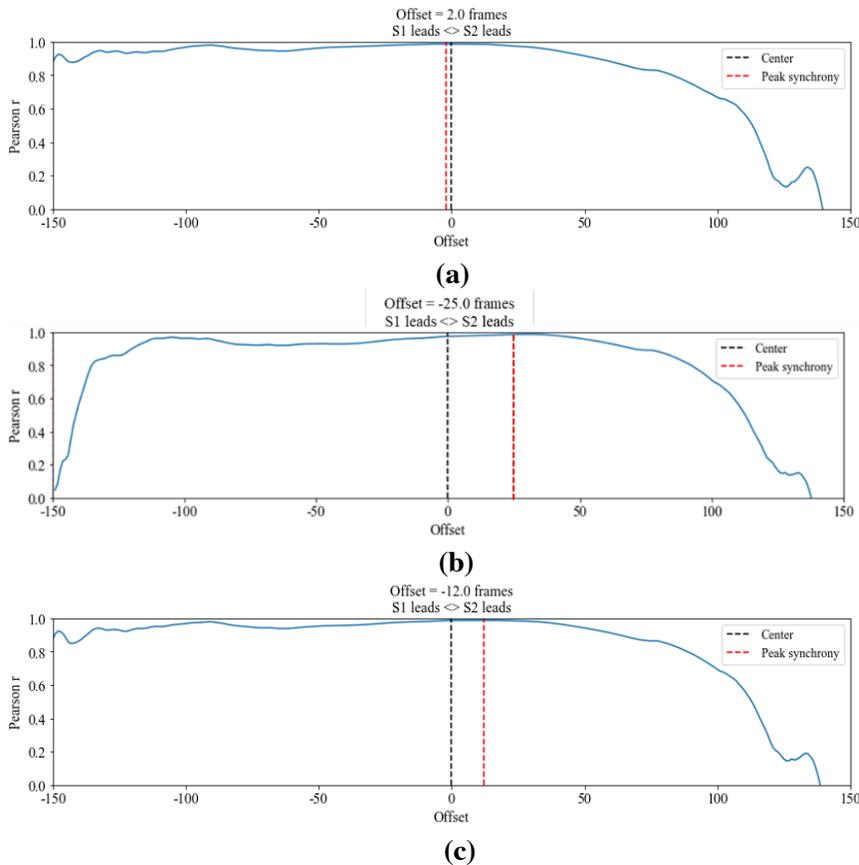

**Figure 10. TLCC result between taxi trip volume and cumulative vaccination data**

## CONCLUSIONS

COVID-19 pandemic has an unprecedented impact on people's daily travel demands. However, with the vaccination rate rising, the travel demand began to recover. Moreover, taxi trips volume, COVID-19 related, and vaccination data provide an abundant and valuable resource to help realize the recovery pattern of taxi demand after vaccination and the factors impacting the recovery of taxi demand. This study firstly investigates the spatial-temporal pattern of Taxi trips in Chicago, including normalized volume and travel distance, before and after vaccination. Secondly, the Pearson coefficient and DTW methods of correlation analysis are utilized to analyze the relationship among pandemic severity, including the daily new and the cumulative number of cases, hospitalization, and death, vaccination progress, including the daily new and cumulative first dose, second dose and total dose vaccination, and taxi trip volume. The factors that most affect the recovery of taxi demand-cumulative vaccination are identified. Furthermore, the leading and lagging relationship between taxi trip volume and cumulative vaccinations are examined to further investigate how the cumulative vaccination impacts the recovery of taxi demand.

After vaccination, taxi trip volume gradually recovers but still needs a long time to reach the pre-pandemic level. In recovery pattern analysis of taxi trips, the result reveals that taxi trip volume increases with the vaccination rate. The most prominent areas of fast recovery are in the airport area and near the central downtown of Chicago. The average travel distance increases in the majority of community areas after vaccination began. Moreover, the recovery of taxi demand is not only related to the pandemic severity but strongly correlated with the cumulative vaccinated population. The result of TLCC found that, after the first dose vaccination, people take taxis more frequently, and the taxi trip volume starts to recover before the second dose vaccination. At the beginning of vaccination, the recovery rate of taxi trip volume is lower than the growth rate of vaccinated people. As the daily taxi trip volume stably increases and more people are vaccinated, the recovery rate of taxi trip demand is larger than the growth rate of the vaccinated population in mid-April 2021. Thence, transportation agencies need to pay attention to increased taxi trip demand before the vaccination rate reached the safe level, such as herb immunity. The government should also concern that mass travel trips by these people who are not fully vaccinated could lead to another pandemic outbreak, which may pose further negative impacts on the economy and society. Therefore, the local government should recommend vaccinating or unvaccinated people to maintain social distancing and keep the nonpharmaceutical intervention (NPIS) in the post-pandemic stage.

There are several limitations of this study worth mentioning. First, multiple factors impact the recovery of taxi demand, such as household income (*34*) and government policy (*35*). This paper only explores the vaccination factors. If more comprehensive data or surveys are obtained, a more comprehensive correlation analysis could be done. Second, the transportation mode in the paper is limited to taxi. Taxi trip data may be biased because people with low income might not take taxies. Therefore, more data on different transportation modes such as rideshare, public transit and bike-share could reduce the bias and make the mobility analysis more comprehensive.

## AUTHOR CONTRIBUTIONS
The authors confirm contribution to the paper as follows: study conception and design: Zihao Li, Yunlong Zhang and Xiaoqiang Kong; literature review: Cheng Zhang and Zihao Li; data collection and preprocessing: Zihao Li; analysis and interpretation of results: Zihao Li, Cheng Zhang, Yunlong Zhang, Xiaoqiang Kong and Chaolun Ma; draft manuscript preparation: Zihao Li, Cheng Zhang, Yunlong Zhang, Xiaoqiang Kong and Chaolun Ma. All authors reviewed the results and approved the final version of the manuscript.